\documentclass[%
prl,
reprint,
superscriptaddress,
showpacs,
showkeys,
nofootinbib,
nobibnotes,
 amsmath,amssymb,
 aps,
]{revtex4-1}

\usepackage{graphicx}% Include figure files
\usepackage[T1]{fontenc} %Extra fonts
\usepackage{epstopdf} %Include eps figures in pdflatex
\usepackage{dcolumn}% Align table columns on decimal point
\usepackage{bm}% bold math
\usepackage{hyperref}% add hypertext capabilities
\usepackage[mathlines]{lineno}% Enable numbering of text and display math
\usepackage{xfrac} %Allows slanted fractions 
\usepackage{color}

\usepackage{amssymb}
\usepackage{amsbsy}
\usepackage{amsmath}
\newcommand{\bp}{$\beta^-\text{p}^+$}

\begin{document}
\preprint{APS/123-QED}

\title{Direct observation of proton emission in $^{11}$Be}

\author{Y.~Ayyad} \email{Corresponding author. ayyadlim@frib.msu.edu}

\affiliation{Facility for Rare Isotope Beams, Michigan State University, East Lansing, Michigan 48824, USA}
\affiliation{National Superconducting Cyclotron Laboratory, Michigan State University, East Lansing, Michigan 48824, USA}

\author {B.~Olaizola}
\affiliation{TRIUMF, 4004 Wesbrook Mall, Vancouver, British Columbia V6T 2A3, Canada}

\author {W.~Mittig}
\affiliation{National Superconducting Cyclotron Laboratory, Michigan State University, East Lansing, Michigan 48824, USA}
\affiliation{Department of Physics and Astronomy, Michigan State University, East Lansing, MI 48824, U.S.A.}

\author{G.~Potel}
\affiliation{Facility for Rare Isotope Beams, Michigan State University, East Lansing, Michigan 48824, USA}

\author{V.~Zelevinsky}
\affiliation{Facility for Rare Isotope Beams, Michigan State University, East Lansing, Michigan 48824, USA}
\affiliation{National Superconducting Cyclotron Laboratory, Michigan State University, East Lansing, Michigan 48824, USA}
\affiliation{Department of Physics and Astronomy, Michigan State University, East Lansing, MI 48824, U.S.A.}
 
\author{M.~Horoi}
\affiliation{Department of Physics, Central Michigan University, Mount Pleasant, MI 48859, USA}

\author {S.~Beceiro-Novo}
\affiliation{Department of Physics and Astronomy, Michigan State University, East Lansing, MI 48824, U.S.A.}

\author {M.~Alcorta}
\affiliation{TRIUMF, 4004 Wesbrook Mall, Vancouver, British Columbia V6T 2A3, Canada}

\author {C.~Andreoiu}
\affiliation{Department of Chemistry, Simon Fraser University, Burnaby, BC V5A 1S6, Canada}

\author {T.~Ahn}
\affiliation{Department of Physics, University of Notre Dame, Notre Dame, Indiana, 46556, USA}

\author {M.~Anholm}
\affiliation{TRIUMF, 4004 Wesbrook Mall, Vancouver, British Columbia V6T 2A3, Canada}
\affiliation{Department of Physics and Astronomy, University of Manitoba, Winnipeg, Manitoba, R3T 2N2, Canada}

\author {L.~Atar}
\affiliation{Department of Physics, University of Guelph, Guelph, Ontario N1G 2W1, Canada}

\author {A.~Babu}
\affiliation{TRIUMF, 4004 Wesbrook Mall, Vancouver, British Columbia V6T 2A3, Canada}

\author {D.~Bazin}
\affiliation{National Superconducting Cyclotron Laboratory, Michigan State University, East Lansing, Michigan 48824, USA}
\affiliation{Department of Physics and Astronomy, Michigan State University, East Lansing, MI 48824, U.S.A.}

\author {N.~Bernier}
\affiliation{TRIUMF, 4004 Wesbrook Mall, Vancouver, British Columbia V6T 2A3, Canada}
\affiliation{Department of Physics and Astronomy, University of British Columbia, Vancouver, BC, V6T 1Z4, Canada}

\author {S.~S.~Bhattacharjee}
\affiliation{TRIUMF, 4004 Wesbrook Mall, Vancouver, British Columbia V6T 2A3, Canada}

\author {M.~Bowry}
\affiliation{TRIUMF, 4004 Wesbrook Mall, Vancouver, British Columbia V6T 2A3, Canada}
\altaffiliation{Present address: School of Engineering, Computing and Physical Sciences, University of the West of Scotland, High Street, Paisley PA1 2BE, United Kingdom}

\author {R.~Caballero-Folch}
\affiliation{TRIUMF, 4004 Wesbrook Mall, Vancouver, British Columbia V6T 2A3, Canada}

\author {M.~Cortesi}
\affiliation{National Superconducting Cyclotron Laboratory, Michigan State University, East Lansing, Michigan 48824, USA}

\author{C.~Dalitz}
\affiliation{Niederrhein University of Applied Sciences, Institute for Pattern Recognition\\ Reinarzstr. 49, 47805 Krefeld, Germany}

\author {E.~Dunling}
\affiliation{TRIUMF, 4004 Wesbrook Mall, Vancouver, British Columbia V6T 2A3, Canada}
\affiliation{Department of Physics, University of York, Heslington, York YO10 5DD, United Kingdom}

\author {A.B.~Garnsworthy}
\affiliation{TRIUMF, 4004 Wesbrook Mall, Vancouver, British Columbia V6T 2A3, Canada}

\author {M.~Holl}
\affiliation{TRIUMF, 4004 Wesbrook Mall, Vancouver, British Columbia V6T 2A3, Canada}
\affiliation{Department of Astronomy and Physics, Saint Marys University, Halifax, Nova Scotia B3H 3C3, Canada}

\author {B.~Kootte}
\affiliation{TRIUMF, 4004 Wesbrook Mall, Vancouver, British Columbia V6T 2A3, Canada}
\affiliation{Department of Physics and Astronomy, University of Manitoba, Winnipeg, Manitoba, R3T 2N2, Canada}

\author {K.~G.~Leach}
\affiliation{Department of Physics, Colorado School of Mines, Golden, Colorado 80401, USA}

\author {J.S.~Randhawa}
\affiliation{National Superconducting Cyclotron Laboratory, Michigan State University, East Lansing, Michigan 48824, USA}

\author {Y.~Saito}
\affiliation{TRIUMF, 4004 Wesbrook Mall, Vancouver, British Columbia V6T 2A3, Canada}
\affiliation{Department of Physics and Astronomy, University of British Columbia, Vancouver, BC, V6T 1Z4, Canada}

\author {C.~Santamaria}
\affiliation{Nuclear Science Division, Lawrence Berkeley National Laboratory, Berkeley, California 94720, USA}

\author{P.~\v{S}iuryt\.{e}}
\affiliation{TRIUMF, 4004 Wesbrook Mall, Vancouver, British Columbia V6T 2A3, Canada}
\affiliation{Department of Physics, University of Surrey, Guildford, Surrey, GU2 7XH, United Kingdom}

\author {C.E.~Svensson}
\affiliation{Department of Physics, University of Guelph, Guelph, Ontario N1G 2W1, Canada}

\author {R.~Umashankar}
\affiliation{TRIUMF, 4004 Wesbrook Mall, Vancouver, British Columbia V6T 2A3, Canada}

\author {N.~Watwood}
\affiliation{National Superconducting Cyclotron Laboratory, Michigan State University, East Lansing, Michigan 48824, USA}

\author {D.~Yates}
\affiliation{TRIUMF, 4004 Wesbrook Mall, Vancouver, British Columbia V6T 2A3, Canada}
\affiliation{Department of Physics and Astronomy, University of British Columbia, Vancouver, BC, V6T 1Z4, Canada}

\date{\today}

\begin{abstract}
The elusive $\beta^-\text{p}^+$ decay was observed in $^{11}$Be by directly measuring the emitted protons and their energy distribution for the first time with the prototype Active Target Time Projection Chamber (pAT-TPC) in an experiment performed at ISAC-TRIUMF. The measured $\beta^-\text{p}^+$ branching ratio is orders of magnitude larger than any previous theoretical model predicted. This can be explained by the presence of a narrow resonance in $^{11}$B above the proton separation energy.
\end{abstract}

%\pacs{
%21.10.-k, %Properties of nuclei; nuclear energy levels
%}

\maketitle

%Nuclei with an excess of neutrons will naturally undergo a $\beta^-$ decay weak process transforming a neutron into a proton ($\text{n} \rightarrow \text{p}^+ + \text{e}^- + \overline{\nu_\text{e}}$). When the $\beta^-$ decay occurs near the so-called neutron-drip line, there is a chance that the resulting daughter nucleus is left in an excited state above the neutron separation energy, and thus a delayed neutron will be emitted (known as $\beta^-$-delayed neutron emission). In a similar way, nuclei with a deficiency of neutrons, will decay through the $\beta^+$ process, with a proton transforming into a neutron and the emission of a positron and a neutrino (or the equivalent process of electron capture ($\epsilon$), in which a proton captures an atomic electron to become a neutron). Likewise, when the decay occurs near the proton drip line, there is a chance of the daughter nuclei emitting a proton, known as $\epsilon/\beta^+$-delayed proton emission. All these processes have in common that they increase the binding energy per nucleon; i.e. it approaches the \textit{valley of stability} and increases the system stability.

It would seem, \textit{a priori}, that $\beta^-$-decay and proton emission are incompatible processes in a neutron-rich nucleus. The $\beta^-$ decay moves isotopes north-west in the nuclear chart, towards the valley of stability and more bound systems, while the proton emission takes it south, becoming a less stable system. This process, called $\beta^-$-delayed proton emission ($\beta^-\text{p}^+$), is energetically forbidden in all but some nuclei for which their neutron separation energy is S$_n<782$~keV~\cite{Baye2011}. Only a handful of $A\leq31$ nuclei that fulfil this condition have been observed, of which the most promising candidate is $^{11}$Be $\rightarrow^{10}$Be + p$^+ + \beta^-$.

Nuclei approaching the nuclear drip lines have large Q$_\beta$ values available and decay into nuclei with low particle binding energy. This opens an energy window where different $\beta$-delayed particle emission channels are allowed~\cite{Borge2013b}. $^{11}$Be, which is the last bound odd beryllium isotope, has several of them energetically available: $\beta\alpha$ (Q$_{\beta\alpha}$=2845.2(2)~keV, $b_\alpha$=3.30(10)\%~\cite{Refsgaard2019}), $\beta$t (Q$_{\beta\text{t}}$=285.7(2) keV, not observed), $\beta^-\text{p}^+$ (Q$_{\beta\text{p}}$=280.7(3) keV, $b_p$=$8.3(9) \cdot 10^{-6}$, indirectly observed~\cite{Riisager2014a}) and $\beta$n (Q$_{\beta\text{n}}$=55.1(5)~keV, not observed).

This low neutron binding energy allows for the counterintuitive $\beta^-\text{p}^+$ decay to happen. In nuclei like $^{11}$Be, called halo nuclei, the last neutron is so weakly bound that it orbits far from an inert core~\cite{Tanihata1996, Al-Khalili2004, Tanihata2013}. The $\beta^-\text{p}^+$ decay mechanism has been modeled as a decay of the halo neutron into a proton that is either in a high-energy resonant state above its separation energy or directly in the continuum, free to be emitted~\cite{Riisager2014b}. This is similar to the $\beta$-delayed deuterium emission in the two-neutron halo nucleus $^{11}$Li. When one of the neutrons decays into a proton, it binds to the other forming a deuteron coupled to the continuum~\cite{Raabe2008}. 

%$^{11}$Be is a one-neutron halo nucleus~\cite{Tanihata1988, Fukuda1991} and the $\beta$-delayed tritium emission is expected to be more suppressed than the $\beta^-\text{p}^+$ branch. When the halo neutron decays into a proton, its wave-function has a small overlap with the $^{10}$Be core to couple to two of its neutrons.

Riisager and collaborators performed experiments at ISOLDE to indirectly observe the $^{11}$Be $\rightarrow^{10}$Be decay~\cite{Borge2013a, Riisager2014a, Riisager2017}. They implanted $^{11}$Be (T$_{1/2}$=13.8~s) in a catcher. Using accelerated mass spectroscopy, they measured the amount of $^{10}$Be (T$_{1/2}=1.5 \cdot 10^6$~years) in the samples, obtaining $b_p=8.3(9) \cdot 10^{-6}$. Theoretical predictions using the $^{11}$B nuclear structure suggest that the $\beta^-\text{p}^+$ branch should have been orders of magnitude lower than observed~\cite{Baye2011,Borge2013a,Riisager2014b}.

%There are two possible explanations for this discrepancy. The $\beta^-\text{p}^+$ decay could populate some unobserved resonance in $^{11}$B, changing the available phase space and enhancing the decay, or there could be additional, unknown decay channels contributing to the $^{11}$Be $\rightarrow^{10}$Be decay. In Ref.~\cite{Riisager2014a} the total branching ratio of the $^{11}$Be $\rightarrow^{10}$Be decay was measured. The $\beta^-\text{p}^+$ and other hypothetical branches, such as the decay to a dark matter particle~\cite{Fornal2018, Pfutzner2018}, were not individually identified. The implications of this alternative dark decay in $^{11}$Be will be discussed at length in the other publication reporting results of this experiment~\cite{Olaizola2019b}.

There are two possible explanations for this discrepancy. The $\beta^-\text{p}^+$ decay could populate some unobserved resonance in $^{11}$B, changing the available phase space and enhancing the decay, or there could be additional, unaccounted for decay channels contributing to the total $^{11}$Be $\rightarrow^{10}$Be decay. Recently, it was suggested in Refs.~\cite{Fornal2018, Pfutzner2018} that the halo neutron in $^{11}$Be is so weakly bound that it could decay into a particle of the dark sector, thus creating a $^{10}$Be nucleus and an undetectable dark particle. The experiment in Ref.~\cite{Riisager2014a} measured the total branching ratio of the $^{11}$Be $\rightarrow^{10}$Be decay, not the individual branching of the $\beta^-\text{p}^+$ and other hypothetical dark decay channels. The implications of this alternative dark decay in $^{11}$Be will be discussed at length in the other publication reporting results of this experiment~\cite{Olaizola2019b}.

The aim of our experiment was to directly measure the $\beta^-\text{p}^+$ decay in $^{11}$Be and unambiguously assign a branching ratio to that specific decay channel. By directly observing the emitted protons and measuring their energy distribution, important information can be extracted about $^{11}$B nuclear structure and the $\beta^-\text{p}^+$ decay mechanism.

In order to efficiently detect and identify every particle emitted in the $\beta$-decay of $^{11}$Be, specially protons of $\sim 200$~keV of energy~\cite{Riisager2014a,Riisager2014b}, the experiment was performed with the prototype Active Target Time Projection Chamber (pAT-TPC)~\cite{SUZUKI201239}. This device allows for efficient and high-resolution measurement of very low energy particles. The pAT-TPC consists of a cylindrical gaseous volume of 50~cm length with 12~cm radius with a detection plane composed of a dual micropattern gas detector (Micromegas~\cite{GIOMATARIS199629} coupled to a Multi-layer Thick Gas Electron Multiplier~\cite{CORTESI2017}). An electric field is applied along the beam axis between the cathode end and the detection plane. Ionization electrons released when a charged particle is crossing the gaseous volume are drifted to the segmented pad plane. There they are multiplied and collected in the Micromegas readout pads. The drift time and the energy loss rate are recorded by each pad individually. With this information and the centroids of the pads, the characteristic energy loss curve of each particle and their tracks can be reconstructed. Each particle track was analyzed independently using a sophisticated clustering algorithm~\cite{DALITZ2019159}. Further information about the pAT-TPC can be found in Refs.~\cite{SUZUKI201239,BRADT201765,Ayyad2019Review}.  

The $^{11}$Be isotopes were produced by a 480-MeV proton beam (9.8~$\mu$A of intensity) delivered by the TRIUMF main cyclotron~\cite{Bylinskii2014} impinging on a uranium carbide (UC$_x$) target. The TRIUMF Resonant Ionization Laser Ion Source~\cite{Lassen06} selectively ionized Be isotopes to a $1^+$ state. After separation of $A/q$=11, particles were further stripped to a $q$=$2^+$ state and re-accelerated to $\sim 390$~A keV. The beam intensity was attenuated to $\sim 10^5$ particles per second to reduce deadtime. The experiment was run in a cycled mode: ions were implanted for 1~s, the charge carriers produced during the implantation were evacuated for 0.5~s and finally 7~s were used to observe $^{11}$Be decay. The beam energy was selected to stop the $^{11}$Be at the center of the pAT-TPC. The ions were expected to neutralize and undergo minimal Brownian motion. However, most of the $^{11}$Be drifted to the cathode where they decayed. The response of the pAT-TPC to low-energy protons was determined by injecting protons in the detector. A molecular OH$^+$ beam was extracted from the Off-Line Ion Source and accelerated before impacting a thin foil to break up the molecule. The p$^+$ was then further accelerated to 220 keV/A. Protons entered the TPC gas volume with $E=198$~keV, as inferred from energy loss parameterizations. The energy resolution of the detector, which amounts to about 15~keV (FWHM) was obtained by integrating the energy loss curve for each proton event to determine the total deposited charge.

The detector was filled with 60~torr of $^{4}$He+CO$_{2}$ (90-10$\%$) gas mixture to stop 200~keV protons within $\sim 10$~cm. Due to the extremely low branching ratio expected for the $\beta^-\text{p}^+$ channel, two different triggers, {\it low-} and {\it high}-level, were used to separate regions of interest in the energy spectrum: 20-300~keV and 300-1500~keV, respectively. The latter was downscaled by a factor of 64 to reduce the triggers of the $\beta\alpha$ decay.

\begin{figure}
  \includegraphics[width=\linewidth]{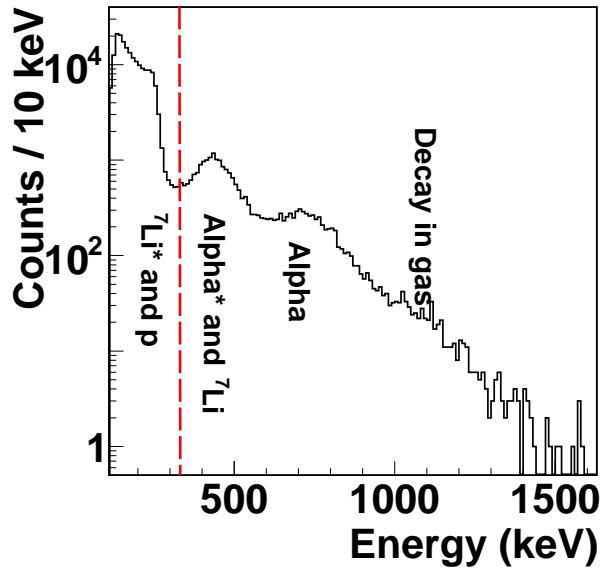}
  \caption{$^{11}$Be $\beta^-$-decay energy spectrum  measured with the pAT-TPC. The dashed line separates both energy regions defined by the {\it low-level} and {\it high-level} triggers. Alpha$^{*}$ refers to the $^{4}$He emitted together with $^{7}$Li*.}
  \label{fig:11bespectrum}
\end{figure}

Figure~\ref{fig:11bespectrum} shows the total energy spectrum of charged particles emitted following the $^{11}$Be $\beta$-decay. The vertical dashed line indicates the separation between trigger windows. One of the products of the decay into $^{7}$Li + $^{4}$He and $^{7}$Li* + $^{4}$He, coming from the cathode of the detector, was identified on an event-by-event basis. In addition, decays in the gas, in which both products were detected simultaneously, ($\sim 10\%$ of the total) were also measured (1200~keV).

%It is important to note that for the highest energy, $^{4}$He have a high probability of punching through the detector depositing less energy. 

\begin{figure}
  \includegraphics[width=\linewidth]{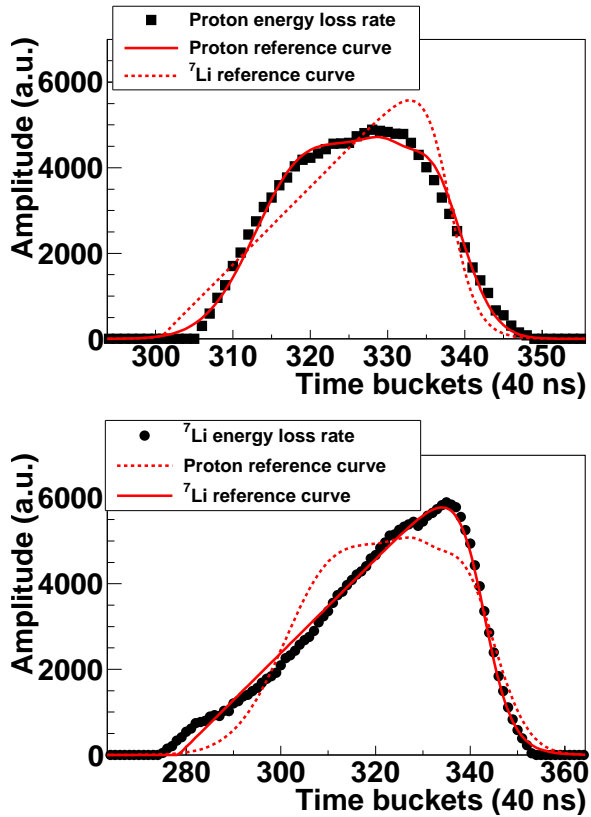}
  \caption{(Upper panel) Experimental energy loss for protons (squares) compared to the reference curves (solid and dotted lines for protons and $^{7}$Li, respectively). The calculated $\chi^{2}$ are 111 (proton reference curve) and 1034 ($^{7}$Li one). (Lower panel) Same as upper panel but for $^{7}$Li* (dotted and solid lines for $^{7}$Li and protons, respectively). In this case, the calculated $\chi^{2}$ is 1890 (proton) and 190 ($^{7}$Li). }
  \label{fig:bragg}
\end{figure}

The identification of protons in the region of interest, which is dominated by $^{7}$Li*, was performed using the characteristic energy loss distribution. An analytical energy loss curve fit to experimental proton and $^7$Li energy loss curves was used to perform a $\chi^{2}$-test for every particle track. These reference curves determined experimentally are shown in Fig.~\ref{fig:bragg} for a $\beta^-\text{p}^+$ decay (upper panel) and a $^{7}$Li*+$^{4}$He decay (lower panel) event. In these events, both the proton and $^{7}$Li* are emitted from the cathode of the detector. The experimental energy loss curves of each event (square for protons and dots for $^{7}$Li*) were compared to both reference curves using an objective function containing the amplitude (energy loss) per unit time (time bucket). Since the reference curves were determined at a fixed angle and energy, they were shaped and re-normalized to take into account the energy-range of the particle, its angle of emission and the starting time bucket of the event. In each step of the algorithm, the objective function is evaluated, selecting the curve that minimizes the $\chi^{2}$.    

The results of the minimization process, in terms of $\chi^{2}$, have also been validated by scrutiny. The $\beta^-\text{p}^+$ decay candidates, 391 in total, were inspected and found to be similar to the one shown in the upper panel of Fig.~\ref{fig:bragg}. 

\begin{figure}
  \includegraphics[width=\linewidth]{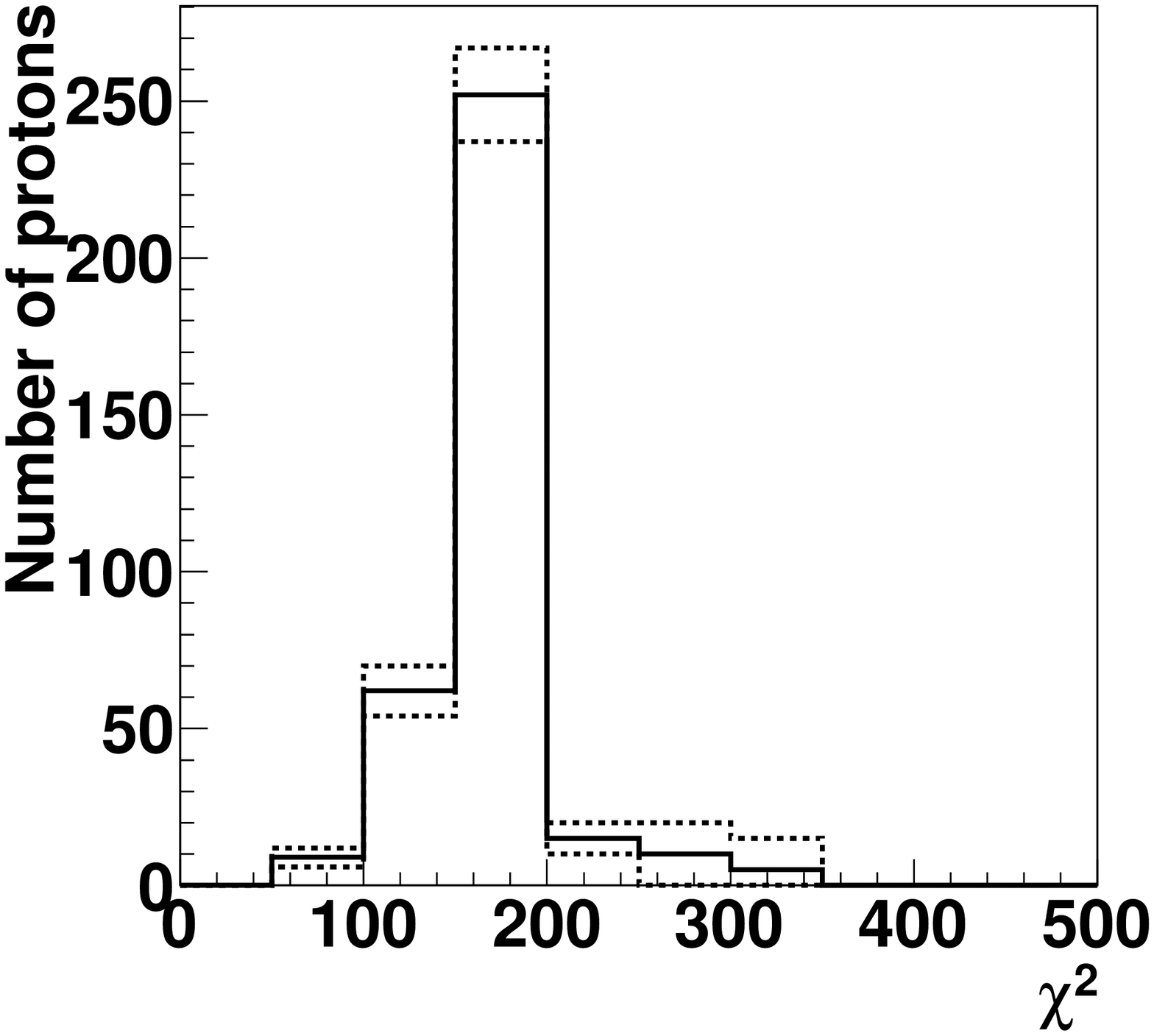}
  \includegraphics[width=\linewidth]{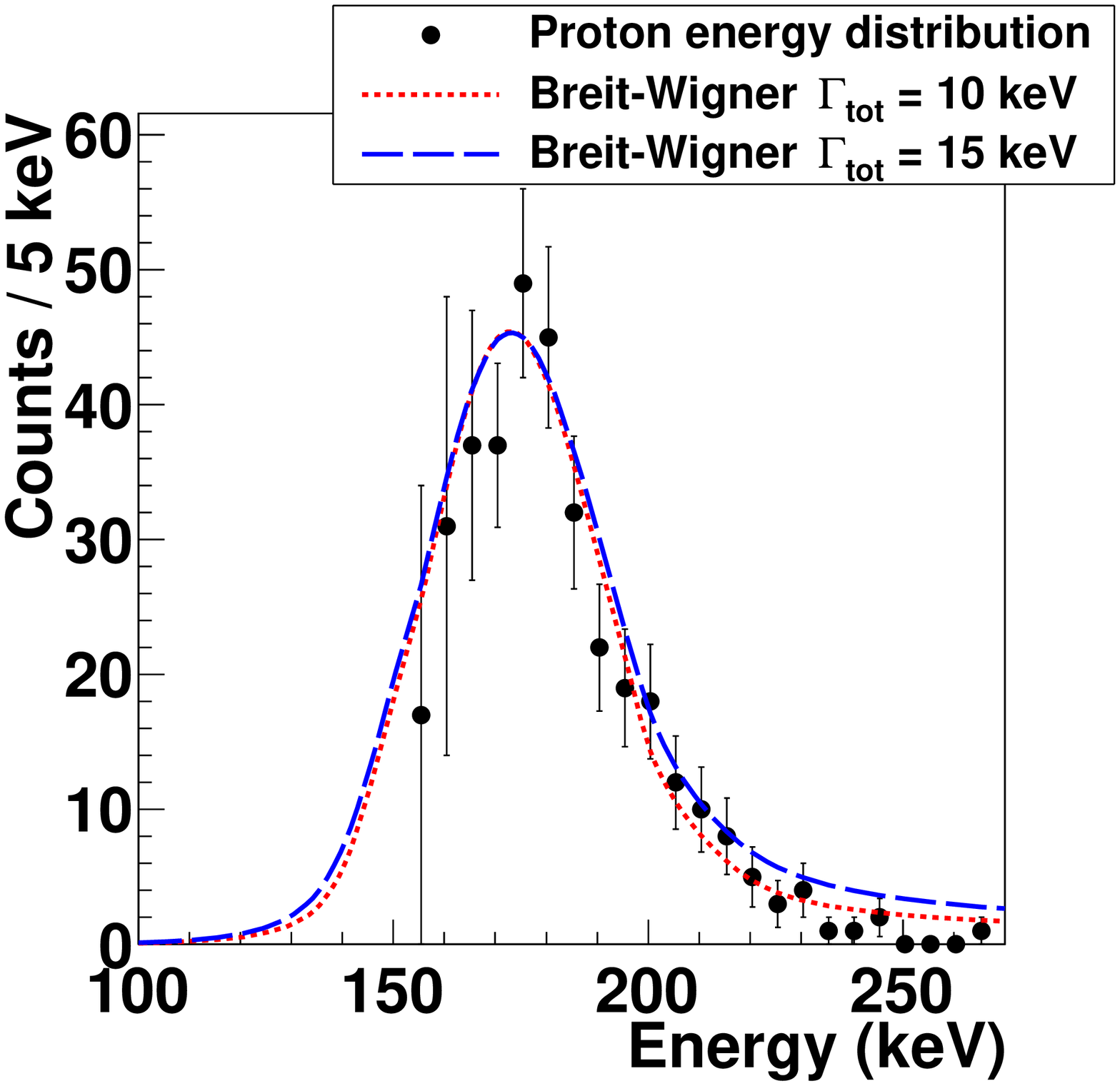}
  \caption{ (Upper panel) Number of identified protons as a function of the $\chi^{2}$ range (solid line). The dashed line refers to the uncertainty limits. (Lower panel) Energy distribution of $\beta$-delayed protons ($\chi^{2}$<200) emitted from $^{11}$Be (solid dots) compared to a Breit-Wigner distribution. }
  \label{fig:energydistribution}
\end{figure}

Background and misidentification of events in the energy spectrum were evaluated by analyzing the distribution as a function of $\chi^{2}$ value (not normalized), as shown in Fig.~\ref{fig:energydistribution}. Below a $\chi^{2}$ value of 200~\footnote{This value was chosen based on a correlation matrix showing the $\chi^{2}$ obtained comparing both reference curves.}, the energy spectrum (solid dots in the lower panel) exhibits a clear peak that is in excellent agreement with a Breit-Wigner distribution with two contributions: protons emitted from the cathode with 178(20)~keV, and 10$\%$ of decays in gas with 196~keV (178~keV plus 18~keV from the $^{10}$Be recoil), respectively, both with a width of 12(5)~keV. The proton penetrability as a function of the energy was also taken into account. The sharp energy distribution clearly indicates that the $\beta^-\text{p}^+$ decay from $^{11}$Be proceeds through an unobserved resonance in $^{11}$B. Based on the Q$_\beta$ and the proton plus recoil energies (the electron screening term is negligible for Be), the resonance is predicted at an energy of 11425(20)~keV. From the proton energy distribution, and taking into account the detector energy resolution, the width of the resonance is $\Gamma = 12(5)$~keV. The branching ratio for the \bp channel was obtained relative to the well known $\beta\alpha$ one, $b_\alpha$=3.30(10)\%~\cite{Refsgaard2019}. The result is $1.3(3) \cdot 10^{-5}$, in agreement with the previous indirect measurement~\cite{Riisager2014a}. The measured branching ratio and resonance energy yields a very small log(\textit{ft})=4.8(4), which suggests a strongly allowed $\beta$ decay. From the $\beta$ decay selection rule of the J$^\pi= 1/2^+$ $^{11}$Be gs.s, we can assume $\Delta J=0,1$ and no change in parity, therefore we propose J$^\pi= (1/2,3/2^+)$ for the newly found resonance.

We evaluated the probability of the direct ``democratic'' decay (simultaneous emission of the proton and $\beta^-$) when there are four particles in the final state: electron (\textbf{p}$_e$, $\epsilon_e$), electron antineutrino (\textbf{p}$_\nu$, $\epsilon_\nu$), proton (\textbf{p}, \textit{E}) and the final nucleus (\textbf{P}, E$\sim 0$).  The release of kinetic energy is 11506-11228=278~keV so that for the electron $p_eR \sim (1/400)(R/\text{fm})$, i.e. always small even for halo nuclei with \textit{R} $\sim 10$~fm (even smaller for the antineutrino). The density of final states for free particles can be estimated, with the normalization to the unit volume, as:

\begin{equation}\label{eq:Vladimir_1}
\begin{split}
 \nonumber    \rho(E) &= (2\pi \hbar)^3 \delta(\textbf{P} + \textbf{p} + \textbf{p}_e + \textbf{p}_\nu)\\
    &\times \delta(\Delta - E - \epsilon_e - \epsilon_\nu)\frac{d^3P d^3p d^3p_e d^3p_\nu}{(2\pi \hbar)^{12}},
\end{split}
\end{equation}

The probability of the decay is:

\begin{equation}\label{eq:Vladimir_3}
\begin{split}
 \nonumber    dw(E)& = \frac{2\pi}{\hbar}G^2|\mathcal{M}|^2 \frac{p^2_\nu do_\nu p^2_e do_e p^2 do_p}{(2\pi\hbar)^9}\\
 &\times dp_e dp_\nu \frac{dp}{dE}dE\delta(X -\epsilon_e - \epsilon_\nu),
 \end{split}
\end{equation}
%\end{widetext}

\noindent where \textit{X} is the total energy of the electron and antineutrino including $m_ec^2$, and $\mathcal{M}$ is the dimensionless sum of the Fermi and Gamow-Teller nuclear matrix elements depending on the halo wave-function.

The integration can be performed exactly:

\begin{equation}\label{eq:Vladimir_4}
    dw(E) =\frac{1}{4\pi^5}\frac{m^5c^4}{\hbar^{10}}G^2|\mathcal{M}|^2 f_0(x)p^2dp,
\end{equation}

where $x = X/mc^2$ and:

\begin{equation}\label{eq:Vladimir_5}
\begin{split}
 \nonumber    f_0(x)& = \frac{1}{5} \left( \frac{1}{6} x^4 - \frac{3}{4} x^2 - \frac{2}{3} \right)\\
 &\times \sqrt{x^2 -1} +\frac{1}{4} x \text{ln} \left( x + \sqrt{x^2-1}  \right).
 \end{split}
\end{equation}

The inverse lifetime follows as (\textit{M} is the nucleon mass)

\begin{equation}\label{eq:Vladimir_6}
   \tau^{-1}= \frac{\sqrt{2}}{4\pi^5} \frac{M^{3/2} m^{13/2} c^7}{\hbar^{10}} G^2 \int^\Omega_0 d\epsilon \sqrt{\epsilon} f_0(x) |\mathcal{M}|^2.
\end{equation}

For calculating the matrix element, we use the analytic form of the neutron halo wave-function~\cite{Hussein1999} and the Coulomb proton wave-function in the field of the charge $Z = 5$. The result for the half-life summing the Fermi and Gamow-Teller contributions is $\tau_{1/2} \sim 2.2 \cdot 10^{10} $~s which corresponds to a branching ratio $6.3 \cdot 10^{-10}$ for the democratic decay, in glaring contrast with the experimental result.

We also estimated the $\beta$-decay rate of $^{11}$Be into a proton-emitting resonance of $^{11}$B, using a simple model to compute the overlap $\mathcal O=\langle\phi_n|\phi_p\rangle$ between the halo neutron $\phi_n(2s_{1/2})$ in $^{11}$Be and the final single-proton resonance $\phi_p(2s_{1/2})$ of $^{11}$B, as well as the width $\Gamma_p$ of the resonance. Both states are calculated making use of Woods-Saxon potentials similar to those used in~\cite{Riisager2014b}, complemented with the Coulomb interaction in the proton case. Setting the resonance energy at the experimental value, the overlap and the width are found to be $\mathcal O=0.53$ and $\Gamma_p=44$~keV, respectively. Within this model, we can also give a rough estimate of the spectroscopic factor $S(^{11}\text{B})$ of the $^{11}$B resonance using the expression $\gamma_p=S(^{11}\text{B})\Gamma_p$, where $\gamma_p$ is the experimental proton emission width. We obtained $S(^{11}\text{B})\approx 0.34$, suggesting that the newly observed $^{11}$B resonance has a sizable single-proton content.

The corresponding partial lifetime $t^p_{1/2}$ can be estimated by integrating the Gamow-Teller strength  over a Lorentzian distribution characterizing the energy spreading of the resonance, with the experimental centroid and width. We assume that the spectroscopic factor of the halo neutron of the $^{11}$Be ground state is $\approx 1$~\cite{Macchiavelli:18}, and that the initial and final isospins are $T_i=3/2$ and $T_f=1/2$. This results in a pure Gamow-Teller transition, with $\langle F\rangle=0$ and  $g_A^2\langle GT \rangle^2=0.15$. With the values of the overlap $\mathcal O$ and the spectroscopic fator $S(^{11}\text{B})$ estimated above, we obtain $t^p_{1/2}=1.7\times10^6~\text{s}$, and $b_p=8.0\times10^{-6}.$

%\begin{equation}\label{eq2}  
%T_{1/2}(E)=\frac{1}{\mathcal %O^2S(^{11}\text{B})}\frac{6141}{g_A^2\langle GT \rangle^2 f(E)},
%\end{equation}

%\noindent where $f(E)$ is the $\beta$-decay phase space, for %substitution in the following:

%\begin{equation}\label{eq3}
%t^p_{1/2}=\frac{\gamma_p}{2\pi}\int_0^{E_{max}} %\frac{T_{1/2}(E)\,dE}{(E-E_{res})^2+\gamma_p^2/4}=1.7\times10^6~\text{s}.
%\end{equation}

 %If we assume that this $^{11}$B resonance decays only through proton %emission, the branching ratio $b_p$ can be obtained from the total %$^{11}$Be lifetime,

%\begin{equation}    \label{eq4}
%b_p=\frac{t_{1/2}(^{11}\text{Be})}{t^p_{1/2}}=8.0\times10^{-6}.
%\end{equation}

%The proton energy distribution measured measured in this experiment is a strong indication that the \bp decay must proceed through a narrow resonance (E=11314(20)~keV, $\Gamma = 12(5)$~keV J$^\pi= (1/2^+)$) in $^{11}$B. The phase space calculations using this unobserved resonance can accurately obtain the measured branching ratio. In contrast, when a democratic decay is considered, the result is orders of magnitude lower than observed.

In summary, we have performed the first direct observation of a $\beta^-\text{p}^+$ decay. This new result is in agreement with the previous indirect measurement~\cite{Riisager2014a}. The use of the pAT-TPC also allowed for a precise measurement of the emitted proton energy, which in turn clarified that the decay proceeds sequentially though a narrow resonance [$E=11425(20)$~keV, $\Gamma = 12(5)$~keV, J$^\pi= (1/2, 3/2^+)$] in $^{11}$B. This resonance, unobserved yet, also explains why the branching ratio is orders of magnitude larger than previous theoretical models predicted. Calculations including this new result can reproduce the measured branching ratio. In contrast, when a democratic decay is considered, the result is orders of magnitude lower than observed. 

\begin{acknowledgments}
The authors would like to thank the beam operators at ISAC-TRIUMF for their effort. They set up a new proton beam of the exact needed energy in a very short time that turned out to be crucial for the successful outcome of the experiment. TRIUMF receives funding through a contribution agreement through the National Research Council Canada. This work was supported by the Natural Sciences and Engineering Research Council of Canada, by the DOE Office of Science (grant DE-SC-0017649, and FRIB Theory Alliance award DE-SC0013617) and the by the U.S. National Science Foundation (NSF) under Cooperative Agreement No. PHY-1565546.
\end{acknowledgments}

\bibliography{bibliography}% Produces the bibliography via BibTeX.

\end{document}